# Optimizing the reliability of a bank with Logistic Regression and Particle Swarm Optimization

**Vadlamani Ravi**[*,1] and **Vadlamani Madhav**[2]

[1]Center of Excellence in Analytics,
Institute for Development and Research in Banking Technology,
Castle Hills Road #1, Masab Tank,
Hyderabad- 500 057, India
vravi@idrbt.ac.in
[2]Department of Electrical Engineering,
Indian Institute of Technology Bombay, Powai, Mumbai-400076, India
vadpadma@gmail.com

**Abstract**

It is well-known that disciplines such as mechanical engineering, electrical engineering, civil engineering, aerospace engineering, chemical engineering and software engineering witnessed successful applications of reliability engineering concepts. However, the concept of reliability in its strict sense is missing in financial services. Therefore, in order to fill this gap, in a first-of-its-kind-study, we define the reliability of a bank/firm in terms of the financial ratios connoting the financial health of the bank to withstand the likelihood of insolvency or bankruptcy. For the purpose of estimating the reliability of a bank, we invoke a statistical and machine learning algorithm namely, logistic regression (LR). Once, the parameters are estimated in the $1^{st}$ stage, we fix them and treat the financial ratios as decision variables. Thus, in the $1^{st}$ stage, we accomplish the hitherto unknown way of estimating the reliability of a bank. Subsequently, in the 2nd stage, in order to maximize the reliability of the bank, we formulate an unconstrained optimization problem in a single-objective environment and solve it using the well-known particle swarm optimization (PSO) algorithm. Thus, in essence, these two stages correspond to predictive and prescriptive analytics respectively. The proposed 2-stage strategy of using them in tandem is beneficial to the decision makers within a bank who can try to achieve the optimal or near-optimal values of the financial ratios in order to maximize the reliability which is tantamount to safeguarding their bank against solvency or bankruptcy.

## 1. Introduction

Reliability theory and engineering found numerous applications in electrical engineering, electronics engineering, mechanical engineering, civil engineering, chemical engineering and software engineering etc. It has, in fact, dramatically changed the way the products are manufactured, electronic devices are designed are designed, civil structures are built and software is tested. Traditionally, reliability is defined as the probability that a component or a system performs its intended function for a specified period of time under stated environmental conditions, where time to failure is treated as a random variable. On the other hand, in reliability physics models, reliability is defined as the probability that the strength of a component or a system is more than the stress exerted on it, where both stress and the strength are treated as random variables. These definitions have been retrofitted with slight modifications to suit the special requirements of various engineering disciplines mentioned above. In almost all of these disciplines, maximization of the reliability is paramount with or without constraints on cost, weight, volume etc.

---

[*] Corresponding author, Phone: +914023294310; FAX: +914023535157

As regards, financial services, especially banks, such a definition of reliability is missing. When one attempts to define the reliability of a bank, three connotations are possible: (i) bank service level reliability (ii) bank network reliability (iii) financial health of the bank characterized by financial ratios which include solvency ratios, liquidity ratios, profitability ratios, etc. While the first facet corresponds to customer service and satisfaction. It is taken care of by effective and holistic implementation of customer relationship management powered by analytics, the second one pertains to inter-bank communication network to facilitate transactions in a shortest path which is ensured by employing proven optimization algorithms.

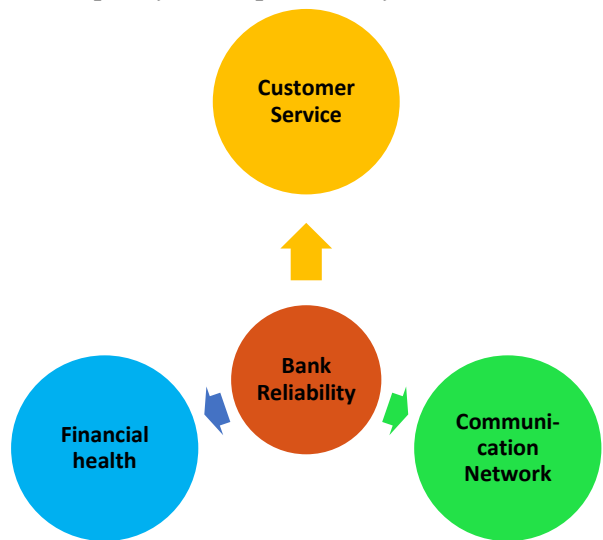

Fig. 1. Different facets of bank reliability

The third facet manifests how healthy a given bank is w.r.t to the liquidity, capital adequacy, efficiency, solvency, profitability etc. Indirectly, it also indicates how the bank is faring so as not to become bankrupt or insolvent beyond the CAMELS rating. This paper is concerned with third facet, where almost little or no work is reported so far.

A bank has many stakeholders viz., customers, bank employees, regulators, investors, auditors etc. When a bank goes bankrupt, all stakeholders are adversely affected. Bankruptcy of banks and firms adversely impacts the economy of any country. Bankruptcy prediction in banks and firms has been pioneered by Altman (1968). Since then, it has been a thriving research area where researchers proposed various techniques ranging from applied statistics to machine learning to soft computing in order to predict bankruptcy as accurately as possible (Ravikumar and Ravi, 2007; Verikas et al, 2009). These survey papers comprehensively reviewed the works reported using statistical, intelligent and soft computing techniques for bankruptcy prediction till 2007 and 2009 respectively.

However, all those techniques without exception are grouped under predictive analytics paradigm because the bankruptcy prediction problem is formulated and solved as a binary classification problem. There, typically a labelled dataset comprising the financial ratios of bankrupt and healthy banks from a specific country is taken and supervised learning techniques are employed in order to solve the underlying binary classification problem. In some cases, when only healthy banks dataset was available, researchers resorted to one-class classification model building, where the characteristics of healthy banks are learnt and they are used to recognise or identify bankrupt banks in future.

However, there was no attempt made so far to prescribe the desirable values of the financial ratios which the bank will try to achieve so as to remain/become healthy or reliable. This paper proposes to precisely fill this gap in literature, whereby (i) the reliability of a bank is defined and estimated for the first time based on the given financial ratios by taking the help of logistic regression (ii) and it is maximized with the help of an optimization technique namely, particle swarm optimization (PSO) in order to obtain the optimal values of the financial ratios along with the maximum reliability. These financial ratios typically include solvency ratios, liquidity ratios, profitability ratios, etc. The current work presents a landmark achievement in bankruptcy research because both predictive analytics and prescriptive analytics are applied in tandem, which is a clear departure from earlier studies. While the 1$^{st}$ stage yields us a definition and an estimate of the bank reliability, the 2$^{nd}$ stage offers prescriptions as to what values of the financial ratios a bank should target in order to achieve maximum reliability thereby remaining solvent or become solvent.

The rest of the paper is organized as follows: Section 2 presents a literature survey of the topic. Section 3 presents the proposed methodology, while section 4 describes the datasets analysed. Section 5 discusses results obtained in the process. Finally, Section 6 concludes the chapter.

## 2. Literature Review

Altman (1968) was the first to study the failure prediction of financial firms and banks in late 1960s. Ever since it has been the most extensively researched area in the last few decades. Since bankruptcy adversely affects creditors, employees, regulators, auditors, stockholders and senior management, all of them interested in bankruptcy prediction (Altman, 1968). Federal Deposit Insurance Corporation Improvement Act of 1991 used a six-part rating system to indicate the safety and soundness of the institution. This rating, called CAMELS rating, evaluates banks and firms in terms of: Capital adequacy, Asset quality, management expertise, Earnings strength, Liquidity, and Sensitivity to market risk. While CAMELS ratings clearly provide regulators with important information, Cole and Gunther (1995) reported that these CAMELS ratings decay rapidly. It is a serious limitation of CAMELS ratings.

Therefore, researchers applied numerous data-driven methods namely logistic regression, discriminant analysis, multi-layer perceptron, radial basis function network, wavelet neural network, probabilistic neural network, group method of data handling network, support vector machine, decision trees, k nearest neighbour, case based reasoning, fuzzy logic based models, rough set theory based models, evolutionary computing, genetic programming etc and their hybrids in various permutations and combinations to solve bankruptcy prediction problem (Ravikumar and Ravi, 2007; Verikas et al, 2009).

In the recent past, some interesting works are published dealing with bankruptcy prediction. Veganzones and Séverin (2018) and Zoricak et al (2020) reported bankruptcy prediction on unbalanced datasets. Hosaka (2019) employed convolutional neural network on the imaged financial ratios data for predicting bankruptcy. Son et al (2019) employed extreme gradient boosting preceded by Box-Cox transformation for predicting bankruptcy and reported interpretable results. Mai et al (2019) reported an innovation in bankruptcy prediction by employing deep learning models on the textual disclosures made by companies. Jardin (2019) proposed to quantize how firm financial health changes over time, typify these changes and design models that fit each type thereby studying the impact of the dynamics of the firm over time on its bankruptcy. Zięba et al (2016) generated synthetic features and ensemble boosted trees for bankruptcy prediction. Liang et al (2016) combined financial ratios and corporate governance indicators to predict bankruptcy. Wang et al (2017) developed grey wolf optimized kernel extreme learning machine for bankruptcy prediction. Volkov et al (2017) predicted bankruptcy of firms using sequential information using Markov classifier. Antunes et al (2017) employed Gaussian process classifier and visualization for bankruptcy prediction. Shi and Li (2019) conducted a bibliometric study of intelligent techniques applied for bankruptcy prediction. Alaka et al (2018) presented a systematic literature review of research reported in bankruptcy prediction and a framework for tool selection. Nyitrai and Virág (2019) studied the impact of outliers on the performance of bankruptcy prediction models. To summarize, all these works have a commonality in that all of them applied various predictive analytics techniques to predict bankruptcy no matter imbalance is present in the datasets or not. Further, none of them prescribed the desirable financial ratio values that a bank should strive for in order to remain solvent or healthy.

Thus, as far as the reliability of a bank is concerned, not much work is reported. Solovjova and Romanova (2020) designed a bank reliability model based on multi-choice ordered logistic regression to assess bank financial strength. The model is developed using publicly available financial statement data of Latvian commercial banks, in the period of 2003–2016, macroeconomic data as well as aggregate statistical data of Latvian banking sector. The model yields ratings indicating the reliability level of banks. The study identifies the most important factors reflecting the reliability level of banks, including shareholder equity ratio, profitability, assets structure, loan portfolio structure, and others. This study come very close to the present study, but deviates significantly by no attempting to define reliability as a probability of non-bankruptcy or solvency and prescribe the desired ratio values in order to remain solvent. They reported that the ratings suggested by their model matched with that of Moody's. Yuan and Gao (2019) surveyed various methods that look into the reliability and validity of the service quality in internet banking. Posnaya et al (2017) suggested a new framework for evaluating reliability of banking system in Russia. They extended the traditional CAMELS framework by considering the additional IT-enabled services, products and operations. They named it as RiCAMELS. Iberahim (2016) investigated the relationship between the reliability and responsiveness of ATM services with customer satisfaction. Data was collected through questionnaire survey of 271 respondents and observations at the service point. Lin et al (2017) studied the service performance of a bank in Taiwan where transaction data is sent from headquarters to bank branches through its transmission system in order to satisfy the customers' requirements. They suggested analysis of the system reliability, which is defined as the probability of demand satisfaction, a key performance indicator for measuring service level of the transmission system. Apart from these, no other work on reliability of bank is reported in literature.

## 3. Proposed Methodology

First, we present a brief overview of the techniques are employed in this paper. Since, logistic regression is too popular to be overviewed, it is briefly described as follows:

The functional form of the logistic regression is defined as follows:

$$y = f(x_1, x_2, \ldots, x_n) = \frac{1}{1+e^{-(\beta_0+\beta_1 x_1+\cdots+\beta_n x_n)}}$$

where $X = (x_1, x_2, \ldots, x_n)$ is the feature vector of financial ratios and where $\beta_0, \beta_1, \ldots, \beta_n$ are the coefficients or the parameters of the logistic regression estimated by Newton-Raphson method or maximum likelihood estimation method.

### 3.1 Particle Swarm Optimization (PSO)

Kennedy and Eberhart (1997) proposed the PSO algorithm which is a population based evolutionary optimization technique. PSO algorithm mimics the behavior of bird flocking, the social behavior of group of people. Each individual particle in PSO is considered to be a point in the N dimensional search space. The procedure of PSO mainly consists of initialization and velocity update. In the initialization phase, randomly generate a population size of solutions called particles with each particle assigned with random velocity ($V_{id}^{Old}$). The neighborhood best ($p_{id}$) is the best path traveled by each of the particle in the population. The global best ($p_{gd}$) is the best path from the entire population. In velocity updation, each particle's velocity is updated with respect to its position ($x_{id}^{old}$) using neighborhood best ($p_{id}$) and global best particle ($p_{gd}$). The velocity ($V_{id}^{New}$) and position ($x_{id}^{New}$) of each particle are updated by using the equations 1 and 2 given below:

$$V_{id}^{New} = w * V_{id}^{Old} + c_1 * rand * (p_{id} - x_{id}^{old}) + c_2 * rand * (p_{gd} - x_{id}^{old}) \qquad (1)$$

$$x_{id}^{New} = x_{id}^{Old} + V_{id}^{New} \qquad (2)$$

Where $c_1$ and $c_2$ are two predefined positive constants (usually $c_1=2$, $c_2=2$ ), w is the inertia weight value, which is continuously decreased as the iterations pass, $rand$ is a random number generated from uniform distribution U(0,1).

### 3.2 Proposed hybrid methodology

Unlike other engineering disciplines, where reliability is defined using either the continuous time domain or stress-strength models, in financial engineering, one cannot immediately think of a counterpart to such definitions.

### 3.2.1 Reliability of a bank

Therefore, a simple but effective definition would be the *"the probability that given bank remains financially healthy and viable at any given period of time given its financial ratios"*.

*$R_X$ (Bank A) = P (Bank A is healthy/ vector of financial ratios X)*

Thus, it implies that in a bank with high reliability, the bank's financial ratios are under control or it achieves their desirable values by taking care of its operations, business and thereby optimally managing credit risk, market risk and operations risk. Because these ratios tend to change over time, by virtue of the bank's business exposure, this definition calls for the reliability assessment of a bank on a periodic basis. Apart from Altman's Z score, no other measure come closer to be considered as an expression for reliability. But Z score is fraught with the disadvantages as follows: (i) it does not attempt to define the probability of health of a bank (ii) it is defined based on the traditional five financial ratios, which were defined by Altman in 1968. However, nature of the banking operations underwent tremendous change since then it calls for incorporating more financial ratios into assessing the bankruptcy of a bank (iii) finally, the coefficients of these ratios, which go into the definition of the Z score are fixed no matter which country they are applied in order to assess the bankruptcy. This is a objectionable as they vary from country to country for obvious reasons.

In absence of any such a measure, data-driven method is needed to define reliability. Therefore, we take recourse to predictive analytics, wherein a labelled dataset of banks from a specific country comprising a set of financial ratios and the corresponding label of a bank being healthy or bankrupt is considered and a binary classifier is built using that dataset. The classifier must output the probability that a given bank is healthy.

Here it throws up several possibilities for the choice of a binary classifier, rendering the operationalization of this definition non-unique. For example, a multi-layer perceptron (MLP), decision tree, support vector machine (SVM) etc. But only those classifiers that output probabilities are suitable. Further, they should be hyperparameter independent and non-parametric. For otherwise, the predicted probabilities will be non-unique. For instance, MLP and probabilistic SVM (PSVM) are candidate methods to define reliability because they output probabilities. But, both MLP and PSVM are unsuitable because they contain hyperparameters to fine tune, rendering the reliability values coming out of them non-unique. Thus, logistic regression exactly fits the bill. We chose logistic regression because it is the simplest nonparametric and reasonably accurate discriminative classifier having no user-defined parameters or hyperparameters to fine tune. This feature makes the probabilities coming out of logistic regression unique. Thus, logistic regression is considered as an approximation for the reliability of the banking system in the specific country with given financial ratios.

*Step-by-step methodology is as follows:*

*1st Stage: Reliability definition and estimation*

1. Build logistic regression model on the entire dataset and extract the estimated parameters of the model. It is virtually the reliability of the banking system under consideration.
2. Save the estimated coefficients or parameters of the reliability.

*They are treated as sacrosanct in the sense that they represent the financial health of the banks under consideration given their financial ratios. The logistic regression model so obtained serves as an estimate for the reliability of the banks in that region. It outputs values between 0 and 1, and hence is suitable for the definition of reliability. This definition takes into consideration all the banks in that region for which the values of the required financial ratios are available. In order to estimate the reliability of any given bank in that country, one needs to substitute its financial ratios into the logistic regression equation. This completes the predictive analytics stage of the proposed methodology.*

### 2nd Stage: Reliability optimization

In this stage, prescriptive analytics is invoked.

3. We fix the parameter values obtained in the 1st stage and formulate an unconstrained optimization problem, where the objective function to be maximized is the reliability, i.e. the logistic function just obtained in the 1st stage with financial ratios being the decision variables.
4. Solve this nonlinear optimization problem with the help of PSO. However, in order to employ PSO, one needs to input the lower and upper bounds on the decision variables. For this purpose, in absence of any other information, we took the lower and upper bounds of each financial ratio in the given dataset and considered them as the bounds for the decision variables.
5. PSO outputs the maximum reliability along with the corresponding optimal solution comprising the financial ratios.

*These optimal financial ratios are the prescriptions given to any bank if it wants to remain healthy. Fig. 2. depicts the schematic of the proposed methodology.*

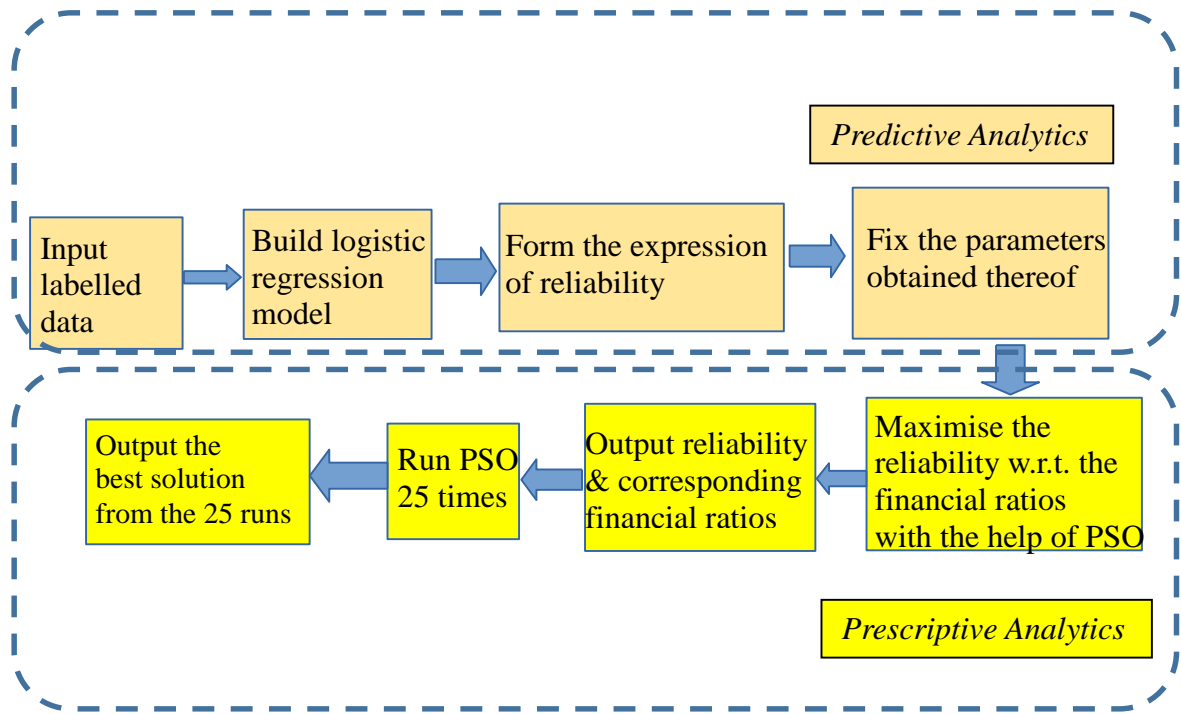

Fig. 2. Schematic of the proposed methodology

## 4. Description of the datasets

We worked on three different bankruptcy datasets viz., Spanish banks, Turkish banks, UK banks. The Spanish banks' data is obtained from Olmeda and Fernandez (1997). Spanish banking industry suffered the worst crisis during 1977-85 resulting in a total cost of 12 billion dollars. The financial ratios are presented in Table 1. The data of the ratios used for the failed banks were taken from the last financial statements before the bankruptcy was declared and the data of non-failed banks was taken from 1982 statements. This dataset contains 66 banks with 9 features, out of which 37 banks went bankrupt and 29 banks are healthy. Turkish banks' data, obtained from Canbas et al. (2005) is available at (http://www.tbb.org.tr/english/bulten/yillik/2000/ratios.xls). Banks association of Turkey published 49 financial ratios. Initially, Canbas et al. (2005) applied univariate analysis of variance (ANOVA) test on these 49 ratios of previous year for predicting the health of the bank in present year. However, only 12 ratios were chosen as the early warning indicators, which have the discriminating ability (significant level is <5%) for healthy and failed banks one year in advance. Among these variables, $12^{th}$ variable has some missing values meaning that the data for some of the banks are not given. So, we filled those missing values with the mean value of the variable following the general approach in data mining. The names of the predictor variables are presented in the Table 1. This dataset contains data of 40 banks and 18 banks went bankrupt and 22 banks are healthy. The UK banks' data is taken from Beynon and Peel (2001). This dataset consists of 60 samples out of which 30 are healthy and 30 are bankrupt.

## 5. Results and discussion

We can see that here the reliability of a bank is quantitatively defined. The effectiveness of the proposed 2-stage method is demonstrated on three different datasets namely Spanish banks, UK banks, and Turkish banks. We used the logistic regression library available in Scikit learn and modified the code for PSO available at https://github.com/nathanrooy/particle-swarm-optimization/blob/master/pso/pso_simple.py .

Table 2. Results of the Banks

| Bank name | Financial ratios | Maximum Iterations | Population size | Next best solutions | Corresponding Reliability |
|---|---|---|---|---|---|
| Spanish | 9 | 3 | 20 | [0.767, 0.494, 0.756, 0.261, 0.02, 0.811, 0.001, 0.474, 0.183] | 0.895 |
|  |  | 5 |  | [0.767, 0.088, 0.807, 0.263, 0.01, 0.799, 0.028, 0.422, -0.023] | 0.896 |
| Turkish | 12 | 5 | 25 | [0.533, 0.627, 0.10, 0.093, 0.509, 0.043, 0.318, 0.543, 0.439, 0.319, 0.867, 0.541] | 0.877 |
|  |  | 3 |  | [0.469, 0.627, 0.409, 0.093, 0.509, 0.043, 0.099, 0.540, 0.275, 0.156, 0.672, 0.654] | 0.845 |
| UK | 12 | 5 | 25 | [15245.342, -19.216, -0.186, 3.210, 1.486, 1.315, 0.418, 0.448, 243.51, 12.091, 1.0, 0.02] | 0.928 |
|  |  | 3 |  | [50688.562, -18.752, -0.328, 3.078, 1.486, 0.497, 1.025, -0.747, 362.935, 44.752, 0.904, 0.122] | 0.915 |

After running the PSO for 25 times in order to nullify the effect of the random seed, we collect the best solution of the population from each run to form an ensemble set of solutions. The next best solutions presented in the Table 2 are taken from that ensemble and are the set of prescribed financial ratios so selected. They have to be adopted by the three regions' banks respectively in order to achieve the near-optimal reliability. We can note that these are the practical prescribed solutions. If the bank is not able to achieve these for some business, technical and operational reasons, values as close as possible to the above sets can be adopted without significantly affecting the reliability by interpolation. This is possible because of the continuous nature of the logistic function.

One can make another interesting observation here in the hindsight. It turns out that the values of the best solution (global optimal solution) vector are indeed the bounds of the financial ratios in the given datasets for most ratios. This arises due to the fact that logistic function is monotonic in its argument. Therefore, optimizing the logistic function is equivalent to optimising the linear function present inside the exponential. Hence, this problem shares a striking similarity with linear programming problem. That is precisely why the vertices of the search space, which is a hyperplane, turned out to be the global optimal solutions in most ratios.

These global optimal solutions, which are easily achieved mathematically owing to the intrinsic linear programming formulation, may not be managerially meaningful or practically achievable. Therefore, we adopt a clever trick. We let the PSO algorithm converge prematurely leading to sub-optimal or near-optimal solutions. The idea of premature convergence is not new to machine learning community because in a neural network like multi-layer perceptron (MLP), we normally resort to premature stopping or convergence in order to avoid over-fitting as indicated by steep rise in the error measure on the test data after certain number of iterations. Thus, we present two such near-optimal solutions in each of the three datasets without significantly sacrificing the maximum achievable reliability. Thus, prescriptive analytics coupled with pragmatic thinking leads to realistic near-optimal reliability together with the corresponding solutions.

This leads to another insightful observation. Since MLP is the composition of several logistic regressions, all of them finally trace back to optimising LR. Therefore, for optimising the reliability of any bank using the definition proposed, optimising the logistic function would be *sufficient*. Further, choosing MLP as a candidate for defining the reliability is fraught with difficulties in the sense that its performance is guided by the choice of its hyperparameters, namely, learning rate, momentum rate and the number of hidden nodes. The same is true with PSVM, where the hyperparameters include regularization value C, choice of the kernel and its associated hyperpararmeters. Thus, different combinations of these hyperparameters lead to different values of the reliability making it non-unique mathematically and eventually resulting in confusion to the decision makers in the bank. This is another reason why MLP or PSVM was not chosen to define and estimate the reliability.

Few other critical remarks are as follows:
- While the definition of reliability is mathematically sound and financially meaningful, its operationalization is limited by the availability of data of the banks in a given country. When data on more banks becomes available, reliability will have to be computed and optimized all over again. This is because logistic regression, just like any statistical or machine learning method tends to yield better predictions (estimates of the reliability) when more data is fed into it.
- Multicollinearity or nonlinear relationships, if at all exists amongst the financial ratios, will have to be accounted for before fitting the logistic regression for estimating the reliability.
- Finally, if in a country, only the data pertaining to the healthy banks is available, which means that either no bank has gone bankrupt so far, or failed banks data is unavailable, then, one has to resort to one-class classification in order to define, estimate and optimize the reliability. The latter two issues are left future work.

## 6. Conclusions

For the first time, we propose a 2-stage hybrid model to define, estimate and optimize the reliability of a bank in a country with respect to its financial ratios from the solvency perspective. The strategy adopted here is to employ predictive and prescriptive analytics in tandem to assure financial health of a banking system. Logistic regression is used here for defining and estimating the reliability of a bank with a given set of financial ratios, while PSO was employed to maximize the reliability with respect to the given financial ratios in an unconstrained framework. The effectiveness of the proposed methodology is demonstrated on three well-known bank datasets. The study recommends the ideal healthy bank scenario, where if the prescribed financial ratios are targeted and achieved, maximum reliability is assured. We noticed that in some cases, the maximum reliability entails financial ratios achieving the lower or upper bounds of the ratios, thereby making the prescription managerially useless even though mathematically meaningful. Therefore, to make the prescriptions more practical, the study prescribes the next best course of action by suggesting near-optimal solutions, which may be mathematically sub-optimal solutions, but are managerially more meaningful and practically achievable.

## Table 1. Financial Ratios of datasets

| SNO. | Financial Ratio Name | |
|---|---|---|
| **Turkish banks' data** | | |
| 1 | Interest Expenses/Average Profitable Assets | IE/APA |
| 2 | Interest Expenses/Average Non-Profitable Assets | IE/ANA |
| 3 | (Share Holders' Equity + Total Income)/(Deposits + Non-Deposit Funds) | (SE+TI)/(D+NF) |
| 4 | Interest Income/Interest Expenses | II+IE |
| 5 | (Share Holders' Equity + Total Income)/Total Assets | (SE+TI)/TA |
| 6 | (Share Holders' Equity + Total Income)/(Total Assets + Contingencies & Commitments) | (SE+TI)/(TA+CC) |
| 7 | Networking Capital/Total Assets | NC/TA |
| 8 | (Salary And Employees' Benefits + Reserve For Retirement)/No. Of Personnel | (SEB+RR)/P |
| 9 | Liquid Assets/(Deposits + Non-Deposit Funds) | LA/(D+NF) |
| 10 | Interest Expenses/Total Expenses | IE/TE |
| 11 | Liquid Assets/Total Assets | LA/TA |
| 12 | Standard Capital Ratio | SCR |
| **Spanish banks' data** | | |
| 1 | Current Assets/Total Assets | CA/TA |
| 2 | Current Assets-Cash/Total Assets | CAC/TA |
| 3 | Current Assets/Loans | CA/L |
| 4 | Reserves/Loans | R/L |
| 5 | Net Income/Total Assets | NI/TA |
| 6 | Net Income/Total Equity Capital | NI/TEC |
| 7 | Net Income/Loans | NI/L |
| 8 | Cost Of Sales/Sales | CS/S |
| 9 | Cash Flow/Loans | CF/L |
| **UK banks' data** | | |
| 1 | Sales | SALES |
| 2 | Profit before tax/capital employed (%) | PBT/C |
| 3 | Funds flow/Total liabilities | FF/TL |
| 4 | (Current liabilities + long term debt)/total assets | (CL/LTD)/TA |
| 5 | Current liabilities/total assets | CL/TA |
| 6 | Current assets/current liabilities | CA/CL |
| 7 | Current assets-stock/Current liabilities | (CA-S)/CL |
| 8 | Current assets-current liabilities/total assets | (CA-CL)/TA |
| 9 | LAG(Number of days between account year end and the date of annual report | LAG |
| 10 | Age | AGE |